# Acoustic Scene Clustering Using Joint Optimization of Deep Embedding Learning and Clustering Iteration

Yanxiong Li, *Member*, *IEEE*, Mingle Liu, Wucheng Wang, Yuhan Zhang and Qianhua He, *Senior Member*, *IEEE*

*Abstract*—Recent efforts have been made on acoustic scene classification in the audio signal processing community. In contrast, few studies have been conducted on acoustic scene clustering, which is a newly emerging problem. Acoustic scene clustering aims at merging the audio recordings of the same class of acoustic scene into a single cluster without using prior information and training classifiers. In this study, we propose a method for acoustic scene clustering that jointly optimizes the procedures of feature learning and clustering iteration. In the proposed method, the learned feature is a deep embedding that is extracted from a deep convolutional neural network (CNN), while the clustering algorithm is the agglomerative hierarchical clustering (AHC). We formulate a unified loss function for integrating and optimizing these two procedures. Various features and methods are compared. The experimental results demonstrate that the proposed method outperforms other unsupervised methods in terms of the normalized mutual information and the clustering accuracy. In addition, the deep embedding outperforms many state-of-the-art features.

## I. Introduction

WITH the prevalence of portable recording devices (e.g., mobile phones) and the development of the mobile internet, the amounts of audio recordings of various classes of acoustic scenes have been rapidly increasing. It is essential to accurately discover classes of acoustic scenes and to efficiently organize massive audio recordings of acoustic scenes for many applications, such as audio surveillance [1], [2], automatic assistance driving [3], and multimedia analysis [4]. Two types of methods are used to perform these tasks: supervised methods (e.g., acoustic scene classification) and unsupervised methods (e.g., acoustic scene clustering). In a supervised method, features are extracted from audio recordings and a pretrained classifier is used to classify each audio recording into one of the predefined classes of acoustic scenes. In an unsupervised method, features are extracted from audio recordings and audio recordings of the same class of acoustic scenes are merged into a single cluster without using prior information and pretrained classifiers.

In this paper, we propose a joint optimization method for acoustic scene clustering. The remainder of the paper is organized as follows. Sections II and III describe related works and our contributions, respectively. Section IV introduces the proposed method. Section V presents the experiments. Finally, the conclusions of this work are presented in Section VI.

## II. Related Works

The related works can be divided into two types: supervised methods and unsupervised methods.

### A. Supervised Methods

Recently, substantial efforts were made on acoustic scene classification in many evaluation campaigns, such as the four editions of the Detection and Classification of Acoustic Scenes and Events (DCASE) challenge [3], [5]-[9]. However, acoustic scene classification was not effectively addressed due to various factors, such as heavy background noise and large variations of time-frequency properties within each class of acoustic scenes [10]. A system for acoustic scene classification typically consists of two key modules: feature extraction and classifier building. The common hand-crafted features include the logarithm mel-band energy, mel frequency cepstral coefficients (MFCCs), spectral flux, spectrogram, Gabor filterbank, cochleograms, I-vector, histogram of gradients features [7], [11], [12], the histogram of gradients [12], hash features [13], and local binary patterns [14], [15]. In addition, some transformed features using matrix factorization [16], [17] and deep neural network [10], [15], [17], [18], are used to address the lack of flexibility of hand-crafted features. The back-end classifiers that were adopted in previous works mainly consist of CNN [19]-[21], gated recurrent neural networks [22], bidirectional long short-term memory networks [10], the Gaussian mixture model, random forest, decision tree, support vector machine, and hidden Markov models [7], [12], [13]. For example, Nguyen et al. proposed a CNN ensemble and nearest neighbor filters for acoustic scene classification [23]. Abeßer et al. classified classes of acoustic scene by combining autoencoder-based dimensionality reduction with a CNN classifier [24]. Eghbal-Zadeh et al. presented an I-vector extraction scheme for acoustic scene classification that uses two audio channels [25]. They used a CNN classifier that was trained on spectrograms of audio excerpts. Valenti et al. used a CNN that was fed by the feature of the logarithm mel spectrogram to classify acoustic scenes [26].

According to the introduction above, most studies have approached acoustic scene classification in a supervised manner. First, they extracted various features (including

This work was supported by the national natural science foundation of China (61771200, 61571192, 6191101570, 6191101514, and 6191101306), the project of international science and technology cooperation of Guangdong province, China (2019A050509001), the open project program of the national laboratory of pattern recognition (201800004), and the fundamental research funds for the central universities, South China University of Technology (Research on key techniques for analyzing complex audio scene contents, 2019).

All authors of this paper are with School of Electronic and Information Engineering, South China University of Technology, Guangzhou, China. The corresponding author is Dr. Yanxiong Li (eeyxli@scut.edu.cn).

time-frequency features and transformed features). Then, they trained a classifier (such as a shallow model or a deep neural network) for each prespecified acoustic scene. Finally, they used the pretrained classifier to determine the acoustic scene class of each test audio recording. In these works, it was assumed that the identities and numbers of acoustic scenes were known in advance. As a result, the main objective of these works was to determine the predefined class of the acoustic scene to which each test audio recording belongs.

*B. Unsupervised Methods*

In practice, we cannot always obtain both the identities and the numbers of classes of acoustic scenes for the following reasons: the ambiguity of acoustic scenes, label loss, weak or incorrect labels, and the high cost of manually labeling massive audio recordings. When we process massive audio recordings, the initial task may be to determine which audio recordings belong to the same class instead of identifying specific acoustic scenes. In this scenario, we need to determine: how to estimate the class number (instead of identities), and how to merge audio recordings of the same class of acoustic scene into a single cluster without using prior information. That is, this problem becomes a problem of acoustic scene clustering.

Very few studies have been conducted on unsupervised methods for relevant types of audio clustering, such as acoustic scene clustering, acoustic event clustering and speaker clustering. For example, Li et al. [27] presented an algorithm for acoustic scene clustering based on a randomly sketched sparse subspace. They used MFCCs as input features of their clustering algorithm, which is an improved spectral clustering algorithm. Li et al. [28] proposed an unsupervised method that was based on the information bottleneck principle for detecting acoustic events. They adopted an agglomerative information bottleneck algorithm for acoustic event clustering. Lu et al. [29] presented an unsupervised method for detecting acoustic events from audio streams. They used a spectral clustering algorithm with context-dependent scaling factors to cluster audio segments. A commonly used unsupervised method for speaker clustering is the AHC algorithm [30]. Theoretically, the unsupervised methods that are discussed above can be applied to acoustic scene clustering. However, they have two main disadvantages. First, hand-crafted and shallow features used in the previous works, cannot effectively represent the property differences among various classes of acoustic scenes. Second, the procedures of feature learning and clustering iteration are sequentially and independently executed, instead of being jointly optimized. In the field of computer vision, a few frameworks are proposed for jointly learning deep representations and image clusters [31]. However, they require the ground-truth number of clusters to be specified in advance and are not suitable for acoustic scene clustering.

III. OUR CONTRIBUTIONS

Inspired by the successes of deep learning in feature representation learning [32] and AHC in data clustering [30], we extract a deep embedding feature and adopt an AHC algorithm for acoustic scene clustering. The deep embedding is learned by a CNN, which is fed by an audio feature of logarithm mel spectrum (LMS). LMS is one of the most popular features for acoustic scene classification [7]. It is used to extract deep embedding. In addition, we propose a framework for jointly optimizing the procedures of both feature learning and clustering iteration by integrating these two procedures into a single model with a unified loss function. In the proposed framework, successive clustering iterations are described as steps in a recurrent procedure, which is piled on a fully connected layer of the CNN for learning the deep embedding. The consideration in the design of the framework is that effective feature representation benefits the clustering performance, while satisfactory clustering results can be used as supervised labels and thus are advantageous for feature representation learning. As a result, more discriminative feature representations and more accurate clustering results can be obtained simultaneously. Fig. 1 presents a schematic diagram that illustrates the difference between our method and traditional unsupervised methods. The deep embedding is used as a feature for clustering in our method.

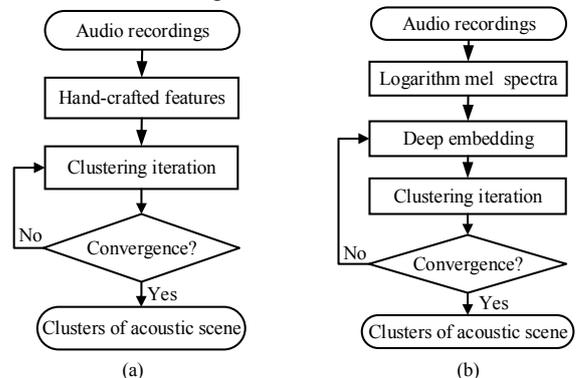

Fig. 1. (a) Traditional unsupervised method and (b) the proposed method.

The main contributions of this study are listed as follows:
1. We propose a recurrent framework for the joint optimization of deep embedding learning and clustering iteration without using manual labels.
2. We formulate a unified loss function for directing the joint optimization.
3. We define an affinity ratio for determining the optimal number of clusters based on a criterion of maximizing the intracluster similarity and minimizing the intercluster similarity.
4. We evaluate the performance of the proposed method and the learned deep embedding under various conditions.

IV. METHOD

In this work, we aim at addressing acoustic scene clustering on untagged audio recordings by jointly optimizing feature representation learning and clustering iteration. We propose a solution that alternates between two procedures: feature representation learning after renewing the network's parameters according to the clustering results of the previous iteration and clustering iteration for updating cluster labels based on the current feature representation. A CNN is used to learn the deep embedding feature, while the AHC algorithm is used to conduct the clustering iterations. The motivation for using CNN to extract the deep embedding feature is based on two considerations. First, CNN performs well in learning feature representations via operations such as convolution and pooling. In addition, the deep embedding feature has been demonstrated to outperform hand-crafted features (such as MFCC) in acoustic event classification [20] and music tagging [33]. Second, CNN has a flexible structure for various tasks and performs well in modeling spectral properties for acoustic scene classification [19], [20], [23]-[26]. The motivation for adopting the AHC algorithm is based on two considerations. First, the AHC algorithm is one of the most predominant algorithms for audio clustering, such as acoustic event clustering [28] and speaker clustering [30]. Second, it is a recurrent procedure and can operate seamlessly in a recurrent framework. In addition, it begins with an initial overclustering and is iteratively merged as satisfactory feature representations are gradually learned by the CNN.

### A. Affinity Definition

Given $N_s$ untagged audio recordings of acoustic scenes $\boldsymbol{S}=\{S_1, \ldots, S_{N_s}\}$ and the initial cluster labels $\boldsymbol{l}=\{l_1, \ldots, l_{N_s}\}$, the CNN parameters $\boldsymbol{\theta}$ are optimized to extract the deep embedding $\boldsymbol{X}=\{x_1, \ldots, x_{N_s}\}$ from $\boldsymbol{S}$. The objective of acoustic scene clustering is to merge $\boldsymbol{X}$ into $N_c$ clusters $\boldsymbol{C}=\{C_1, \ldots, C_i, \ldots, C_{N_c}\}$, where $C_i=\{x_m|\ l_m=i;\ \forall m \in 1, \ldots, N_s;\ \forall i \in 1, \ldots, N_c\}$.

We introduce the graph degree linkage for measuring the affinity (similarity) between two clusters (or audio recordings), which is robust to noise and outliers and is easy to implement [34]. In addition, it has been successfully applied to many tasks, such as image clustering and segmentation [34]. A directed graph $G=(V, E)$ is constructed, where $V$ denotes the set of vertices that correspond to deep embedding $\boldsymbol{X}$ and $E$ denotes the set of edges that connect vertices. An affinity matrix $\boldsymbol{W} \in \mathbb{R}^{N_s \times N_s}$ is defined by Eq. (1), which corresponds to the edge set.

$$w_{m,n} = \begin{cases} \exp\left(-\frac{\|x_m - x_n\|_2^2}{\sigma^2}\right), & \text{if } x_n \in \mathbb{N}_m^{K_s} \\ 0, & \text{otherwise} \end{cases} \quad (1)$$

where $1 \leq m \leq N_s$, $1 \leq n \leq N_s$; $w_{m,n}$ is an element of $\boldsymbol{W}$ and the weight of the edge from vertex $x_m$ to $x_n$; $K_s$ is an integer whose settings will be discussed in the section on the experiments; $\mathbb{N}_m^{K_s}$ denotes the $K_s$ nearest neighbors of $x_m$; and $\sigma^2$ is defined by

$$\sigma^2 = \frac{1}{N_s K_s} \sum_{x_m \in X} \sum_{x_n \in \mathbb{N}_m^{K_s}} \|x_m - x_n\|_2^2 \quad . \quad (2)$$

The affinity between clusters $C_i$ and $C_j$ is defined by

$$A_{C_i,C_j} = \frac{1}{|C_i|^2} \mathbf{1}_{|C_i|}^T \boldsymbol{W}_{C_i,C_j} \boldsymbol{W}_{C_j,C_i} \mathbf{1}_{|C_i|} + \frac{1}{|C_j|^2} \mathbf{1}_{|C_j|}^T \boldsymbol{W}_{C_j,C_i} \boldsymbol{W}_{C_i,C_j} \mathbf{1}_{|C_j|} , (3)$$

where $\boldsymbol{W}_{C_i,C_j}$, which denotes the weights of the edges from clusters $C_i$ to $C_j$, is a submatrix of $\boldsymbol{W}$ whose row indices correspond to the vertices in $C_i$ and whose column indices correspond to the vertices in $C_j$; $|C|$ denotes the cardinality of cluster $C$; $\mathbf{1}_{|C|}$ denotes a vector of all ones of length $|C|$; and T denotes the transpose operation of a matrix (vector).

### B. Joint Optimization Framework

A superscript $t$ is appended to $\{\boldsymbol{X}, \boldsymbol{C}, \boldsymbol{l}, \boldsymbol{\theta}\}$ to refer to the values at timestep $t$, with $1 \leq t \leq T$, where $T$ is total number of iterations and $T < N_s$. It is assumed that a cluster $C_i^{t-1}$ and its nearest cluster $C_j^{t-1}$ are merged from $\boldsymbol{l}^{t-1}$ to $\boldsymbol{l}^t$. Then, a loss function for jointly optimizing $\boldsymbol{X}$ and $\boldsymbol{l}$ at timestep $t$ is defined by a negative affinity

$$\zeta^t(\boldsymbol{l}^t, \boldsymbol{X}^t | \boldsymbol{l}^{t-1}, \boldsymbol{S}) = -A_{C_i^{t-1}, C_j^{t-1}} \quad , \quad (4)$$

where $A_{C_i^{t-1}, C_j^{t-1}}$ is the affinity between cluster $C_i^{t-1}$ and cluster $C_j^{t-1}$, which is defined by Eq. (3); $\zeta()$ is the loss function; $\boldsymbol{l}^0$ denotes the initial cluster labels; and LMS of each audio recording is regarded as an initial cluster. A CNN's parameters $\boldsymbol{\theta}^0$ are initialized for extracting deep embedding $\boldsymbol{X}^0$ based on the LMS of each audio recording. Hence, the cumulative loss of all timesteps is computed by

$$\zeta(\{\boldsymbol{l}^1,\ldots,\boldsymbol{l}^T\},\{\boldsymbol{X}^1,\ldots,\boldsymbol{X}^T\} | \boldsymbol{S}) = \sum_{t=1}^T \zeta^t(\boldsymbol{l}^t, \boldsymbol{X}^t | \boldsymbol{l}^{t-1}, \boldsymbol{S}) \quad . \quad (5)$$

It is difficult to simultaneously obtain optimal clustering results $\{\boldsymbol{l}^1, \ldots, \boldsymbol{l}^T\}$ and deep embeddings $\{\boldsymbol{X}^1, \ldots, \boldsymbol{X}^T\}$ that can minimize the overall loss, which is defined by Eq. (5). Hence, we iteratively obtain the minimum loss value via a recurrent procedure. We fix one element in $\{\boldsymbol{l}, \boldsymbol{X}\}$ and search for the optimal value of the other element. The loss function in Eq. (5) can be decomposed into two alternant steps:

$$\zeta(\{\boldsymbol{l}^1,\ldots,\boldsymbol{l}^T\} | \{\boldsymbol{X}^1,\ldots,\boldsymbol{X}^T\}, \boldsymbol{S}) = \sum_{t=1}^T \zeta^t(\boldsymbol{l}^t | \boldsymbol{X}^t, \boldsymbol{l}^{t-1}, \boldsymbol{S}) \quad , (6)$$

$$\zeta(\{\boldsymbol{X}^1,\ldots,\boldsymbol{X}^T\} | \{\boldsymbol{l}^1,\ldots,\boldsymbol{l}^T\}, \boldsymbol{S}) = \sum_{t=1}^T \zeta^t(\boldsymbol{X}^t | \boldsymbol{l}^{t-1}, \boldsymbol{S}) \quad . \quad (7)$$

Eq. (6) can be regarded as a common clustering problem whose feature representations are specified in advance. Eq. (7) can be regarded as a typical problem of feature representation learning that is supervised by cluster labels.

Fig. 2 illustrates the proposed framework for jointly optimizing both deep embedding learning (by updating the CNN parameters) and clustering iteration (by the AHC algorithm). LMS is extracted from audio recordings. LMS is fed into the CNN at each timestep $t$. Then, the deep embeddings $\boldsymbol{X}^t$ are extracted from the CNN after updating the CNN's parameters from $\boldsymbol{\theta}^{t-1}$ to $\boldsymbol{\theta}^t$ under the supervision of cluster labels $\boldsymbol{l}^{t-1}$. Finally, clustering iteration of the AHC algorithm is conducted on $\boldsymbol{X}^t$ to obtain new clusters $\boldsymbol{C}^t$ and labels $\boldsymbol{l}^t$ based on old clusters $\boldsymbol{C}^{t-1}$ and labels $\boldsymbol{l}^{t-1}$. The alternation between clustering iteration and deep embedding learning is iteratively conducted until the convergence condition is satisfied.

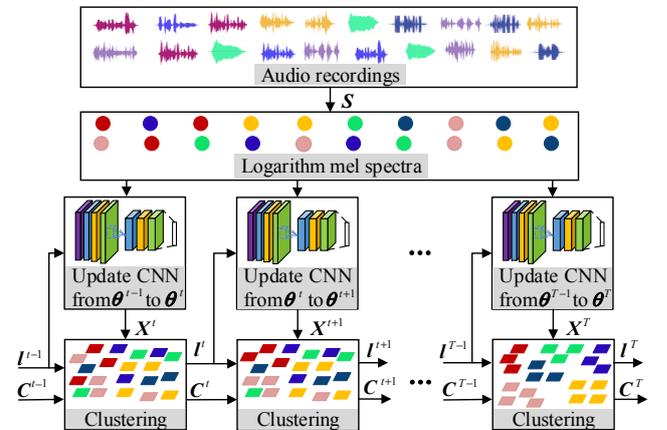

Fig. 2. Proposed framework for jointly optimizing feature representation learning and clustering iteration.

The proposed method is summarized as Algorithm I in Table I. Its steps, namely, LMS extraction, deep embedding learning, clustering iteration and convergence condition, will be described in the following subsections.

TABLE I
ALGORITHM I: PROPOSED JOINT OPTIMIZATION METHOD

**Input**:
  LMS of $N_s$ untagged audio recordings.
**Do**:
  Initialization: $t \leftarrow 0$; the LMS of each audio recording is regarded as an initial cluster and initial cluster label is $l^0$; initializing a CNN parameters $\theta^0$ to extract deep embedding $X^0$.
  **Repeat:**
  $t \leftarrow t+1$.
  Update CNN parameters from $\theta^{t-1}$ to $\theta^t$.
  Extract $X^t$ from the updated CNN parameters $\theta^t$.
  Update clusters from $l^{t-1}$ ($C^{t-1}$) to $l^t$ ($C^t$) by merging two nearest clusters.
  **Until** Convergence condition is met.
**Output**:
  Optimized results: $l^*$ ($C^*$) $\leftarrow l^t$ ($C^t$) and $X^* \leftarrow X^t$.

## C. LMS Extraction

The extraction procedure of LMS is illustrated in Fig. 3. The audio recordings are divided into overlapping audio frames and windowed using a Hamming window [35]. Next, the fast Fourier transform (FFT) is applied on the windowed audio frames to obtain the power spectrum, which is smoothed using a bank of triangular filters. The center frequencies of the filters are uniformly spaced on the Mel-scale.

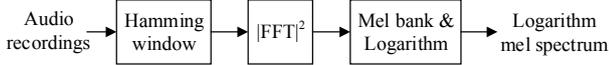

Fig. 3. Extraction procedure of the logarithm mel spectrum.

## D. Deep Embedding Learning

To extract a deep embedding with higher discriminability, we derive the optimal CNN parameters $\theta$ for minimizing the loss that is generated by the clustering iteration. The clustering at timestep $t$ is based on the clustering results of all previous timesteps. The cumulative loss of all timesteps is computed by

$$\zeta(X \mid \{l^1,...,l^t\}, S) = \sum_{k=1}^{t} \zeta^k(X \mid l^{k-1}, S) \quad . \quad (8)$$

Minimizing Eq. (8) with respect to $X$ updates the CNN parameters $\theta$ on LMS under the supervision of $\{l^1, ..., l^{t-1}\}$. The back-propagation algorithm [36] is used to update the CNN parameters from $\theta^{t-1}$ to $\theta^t$ at timestep $t$.

The features that are learned at layers of a CNN can represent the properties of audio recordings. Fig. 4 illustrates an example of the extraction of a deep embedding from a fully connected layer. Convolution layers and fully connected layers are used to conduct the feature mapping from a layer to its next layer.

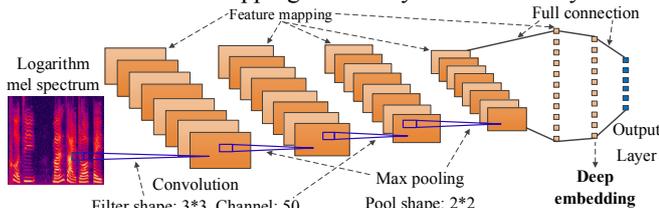

Fig. 4. Illustration of the extraction of deep embedding from a CNN.

## E. Clustering Iteration

After updating the CNN parameters from $\theta^{t-1}$ to $\theta^t$, we can extract deep embeddings $X^t$ at timestep $t$. Then, the deep embeddings in clusters $C^{t-1}$ are updated from $X^{t-1}$ to $X^t$ for the clustering iteration. Finally, two nearest clusters are merged based on clusters $C^{t-1}$ and labels $l^{t-1}$ by searching the maximal affinity over all pairs of clusters. The clustering iteration from clusters $C^{t-1}$ to $C^t$ is summarized as Algorithm II in Table II.

TABLE II
ALGORITHM II: CLUSTERING ITERATION FROM CLUSTERS $C^{t-1}$ TO $C^t$

**Input**:
  Clusters $C^{t-1}=\{C_1, ..., C_{N_C^{t-1}}\}$ and labels $l^{t-1}$ at timestep $t-1$, where deep embeddings in $C^{t-1}$ have been updated from $X^{t-1}$ to $X^t$ already.
**Do**:
  Build $K_s$-nearest neighbor graph, and obtain affinity matrix $W$ by Eq. (1).
  Find two clusters $C_i$ and $C_j$, such that $\{C_i, C_j\}= \arg\max_{C_i, C_j \in C^{t-1}} A_{C_i, C_j}$.
  Merge clusters $C_i$ and $C_j$: $C^t = \{C^{t-1} \setminus \{C_i, C_j\}\} \cup \{C_i, C_j\}$, $N_C^t = N_C^{t-1} -1$.
**Output**:
  Clusters $C^t=\{C_1, ..., C_{N_C^t}\}$ and labels $l^t$ at timestep $t$.

According to the descriptions, we merge one cluster and its nearest neighbor at each timestep such that the loss $\zeta^t(l^t \mid X^t, l^{t-1}, S)$ is minimized over all possible cluster pairs.

## F. Convergence Condition

For the AHC algorithm that is used in this study, the clustering iteration should be terminated when the estimated number of clusters is equal to the real (ground-truth) number of clusters. However, the real number of clusters is typically unknown in practice. Hence, we define an affinity ratio for determining the optimal number of clusters $N_c^*$. In the optimal clustering result, the similarities of audio recordings within each cluster (intracluster affinities) are maximized while the similarities of audio recordings between all pairs of clusters (intercluster affinities) are minimized. Based on this criterion, we define two affinities:

$$A_{intra}^t = \frac{1}{N_c^t} \sum_{i=1}^{N_c^t} A_{C_i^t, C_i^t} \quad, t=1, ..., T, \quad (9)$$

$$A_{inter}^t = \frac{1}{0.5 N_c^t (N_c^t - 1)} \sum_{i=1}^{N_c^t} \sum_{j=i+1}^{N_c^t} A_{C_i^t, C_j^t} \quad, t=1, ..., T, \quad (10)$$

where $A_{intra}^t$ and $A_{inter}^t$ denote the intracluster and intercluster affinities, respectively, at timestep $t$ and $A_{C_i^t, C_j^t}$ (or $A_{C_i^t, C_i^t}$) denotes the affinity between clusters $C_i^t$ and $C_j^t$ (or $C_i^t$), which is defined by Eq. (3). The affinity ratio $\gamma^t$ is defined by

$$\gamma^t = \frac{A_{intra}^t}{A_{inter}^t} \quad, \quad t=1, ..., T. \quad (11)$$

In addition, to reduce the computation time for calculating $\gamma^t$, we set an interval from $N_c^{min}$ to $N_c^{max}$ for determining $N_c^*$. Then, convergence condition is defined as follows:

$$\begin{cases} N_c^* = N_c^\tau, & N_c^{min} \leq N_c^\tau \leq N_c^{max} \\ \tau = \arg\max_{1 \leq t \leq T} \gamma^t \end{cases}, \quad (12)$$

where $1 \leq N_c^{\min} < N_c^{\max} \leq N_s$. Once the clustering algorithm has converged, $N_c^*$ is estimated.

## V. EXPERIMENTS AND DISCUSSIONS

This section introduces two corpora of acoustic scenes and describes the experimental setups. Then, the proposed method is compared with state-of-the-art methods and is evaluated under special conditions.

### A. Experimental Corpora

Experiments are conducted on two corpora of acoustic scenes. The first corpus is the LITIS Rouen audio scene dataset [17], which is denoted as LITIS-Rouen. The second corpus is the dataset of task 1 (acoustic scene classification) of DCASE challenge 2017 [3], which is denoted as DCASE-2017. The corpora DCASE-2017 [1] and LITIS-Rouen [2] are publicly available for research purposes. They are two of the most commonly used experimental datasets in previous studies for acoustic scene classification.

The audio recordings in the LITIS-Rouen corpus were acquired by a Galaxy S3 smart phone that was equipped with Android via the Hi-Q MP3 recorder application. The sampling frequency was 44.1 kHz and the audio recordings were saved as MP3 files with a bit-rate of 705.6 kbps. When transformed into raw audio data, they are downsampled to 22.05 kHz. In this dataset, there are 6052 samples, which are 25.22 hours in length and contain 19 classes of acoustic scenes. Each sample of audio recordings is 15 seconds in length, which was obtained by splitting a long audio recording into 15-second segments without overlaps. Detailed information on the LITIS-Rouen corpus is presented in Table III.

TABLE III
DETAILED INFORMATION ON THE LITIS-ROUEN CORPUS

| Class | NAR | TD(m) | Class | NAR | TD(m) |
|---|---|---|---|---|---|
| Billiard pool hall | 310 | 77.5 | Plane | 46 | 11.5 |
| Quiet street | 180 | 45 | Busy street | 286 | 71.5 |
| Student hall | 176 | 44 | Bus | 384 | 96 |
| Metro-Paris | 278 | 69.5 | Café | 240 | 60 |
| Pedestrian street | 244 | 61 | Car | 486 | 121.5 |
| Train station hall | 538 | 134.5 | shop | 406 | 101.5 |
| Kid game hall | 290 | 72.5 | Train | 328 | 82 |
| High-speed train | 294 | 73.5 | Market | 552 | 138 |
| Tube station | 250 | 62.5 | Restaurant | 266 | 66.5 |
| Metro-Rouen | 498 | 124.5 | **Total** | **6052** | **1513** |

NAR: Number of Audio Recordings; TD(m): Total Duration (minutes).

The DCASE-2017 corpus includes 15 classes of acoustic scenes with distinct recording locations. For each recording location, an audio recording that is 3-5 minutes long is captured. The original audio recordings are split into segments of length 10 seconds. These audio segments are provided in individual files. The audio recordings are obtained using a Soundman OKM II Klassik/studio A3, an electret binaural microphone and a Roland Edirol R-09 wave recorder with a 44.1-kHz sampling frequency and 24-bit resolution. The audio recordings are saved as WAV files with two channels. When transformed into raw audio data, they are downsampled to 22.05 kHz, and the data of only one channel are used in the experiments. The DCASE-2017 corpus contains 5130 samples and is 14.25 hours in length, whose detailed information is presented in Table IV.

[1] http://www.cs.tut.fi/sgn/arg/dcase2017/challenge/task-acoustic-scene-classification
[2] https://sites.google.com/site/alainrakotomamonjy/home/audio-scene

TABLE IV
DETAILED INFORMATION OF THE DCASE-2017 CORPUS

| Class | NAR | TD(m) | Class | NAR | TD(m) |
|---|---|---|---|---|---|
| Lakeside beach | 342 | 57 | Inside bus | 342 | 57 |
| Café/restaurant | 342 | 57 | Library | 342 | 57 |
| Residential area | 342 | 57 | Home | 342 | 57 |
| Grocery store | 342 | 57 | Office | 342 | 57 |
| Metro station | 342 | 57 | Park | 342 | 57 |
| Forest path | 342 | 57 | Train | 342 | 57 |
| City center | 342 | 57 | Tram | 342 | 57 |
| Inside car | 342 | 57 | **Total** | **5130** | **855** |

### B. Experimental Setup

The experiments are implemented on a computer with an Intel(R) Core(TM) i7-6700, 3.10 GHz CPU, 48 GB RAM, and a NVIDIA 1080 TI GPU. Two metrics, namely, the normalized mutual information (*NMI*) and the clustering accuracy (*CA*), are used to measure the performances of various methods since they are widely used as performance metrics [37], [38].

Let $n_{ij}$ be the total number of audio recordings in cluster $i$ that belong to acoustic scene $j$; let $N_g$ be the total number of classes of acoustic scenes (the ground-truth number of clusters); let $N_c$ be the total number of clusters (the estimated number of clusters); let $N_s$ be the total number of audio recordings; let $n_{\cdot j}$ be the total number of audio recordings of acoustic scene $j$; and let $n_{i\cdot}$ be the total number of audio recordings in cluster $i$. The equations in Eq. (13) establish relationships among the above variables:

$$n_{i\cdot} = \sum_{j=1}^{N_g} n_{ij} \ , \ n_{\cdot j} = \sum_{i=1}^{N_c} n_{ij} \ , \ N_s = \sum_{i=1}^{N_c}\sum_{j=1}^{N_g} n_{ij} \ . \quad (13)$$

*NMI* and *CA* are used to measure the agreement between the produced clusters and the ground-truth classes. *NMI* is defined by

$$NMI = \frac{\sum_{i=1}^{N_c}\sum_{j=1}^{N_g} n_{ij} \log\left(\frac{N_s \times n_{ij}}{n_{i\cdot} \times n_{\cdot j}}\right)}{\sqrt{\left(\sum_i n_{i\cdot} \log\frac{n_{i\cdot}}{N_s}\right)\left(\sum_j n_{\cdot j} \log\frac{n_{\cdot j}}{N_s}\right)}} \ . \quad (14)$$

If the clustering results perfectly match the true labels, *NMI* is equal to 1. In contrast, *NMI* approaches 0 if the audio recordings are randomly partitioned. *CA* is defined by

$$CA = \frac{1}{N_s}\left[\sum_{k=1}^{N_s} \delta(y_k, \text{map}(c_k))\right] \ , \quad (15)$$

where $y_k$ and $c_k$ denote the true label and the predicted cluster label, respectively, of the $k^{th}$ audio recording. $\delta(y, c)$ is equal to 1 if $y = c$ and 0 otherwise. map($\cdot$) is a permutation function that maps each cluster label to a true label and an optimal matching can be obtained via the Hungarian algorithm [39]. *NMI* is an information-theoretic measure of the clustering quality. *CA* measures the clustering quality based on the permutation mapping between the true labels and the predicted labels. The higher their values are, the higher the clustering quality is.

The main parameters of the proposed method are optimally tuned on experimental data and their settings are listed in Table V. The settings of integer $K_s$ directly affect the clustering results and will be tuned in next subsection to obtain better results. The experiments are successively conducted on one of the two corpora. The audio recordings of each corpus are evenly divided into two subsets. The first and second subsets are used as development and test data, respectively.

TABLE V
MAIN PARAMETER SETTINGS OF THE PROPOSED METHOD

| Type | Parameters settings |
|---|---|
| Preprocessing | Frame length/overlapping: 40/20 ms<br>Dimension of log mel spectrum: 208 |
| CNN | Batch size: 60<br>Channel number of input layer: 4<br>Learning rate: 0.001<br>Maximum iteration: 3000<br>Weight decay: 0.1<br>Unrolled step: 10<br>Convolution kernel size: 3*3<br>Channel number of convolution layer (CL): 5<br>Number of CL: 2<br>Number of fully connected layer (FCL): 2<br>Neuron numbers of 2 FCLs: (10500, 160)<br>Dimension of deep embedding: 160 |
| AHC | $N_c^{\min}=3$, $N_c^{\max}=30$ |

### C. Estimation of the Number of Clusters

The value of integer $K_s$ influences the values of $w_{m,n}$ in Eq. (1) and $A_{C_i,C_j}$ in Eq. (3) and, thus, affects the estimated optimal number of clusters $N_c^*$ in Eq. (12). In this subsection, we discuss the impacts of $K_s$ on the estimated value of $N_c^*$ and the impacts of the estimated value of $N_c^*$ on the results of acoustic scene clustering. In this experiment, our method is evaluated on the development data of LITIS-Rouen and DCASE-2017. As shown in Fig. 5, the estimated number of clusters $N_c^*$ changes with the integer $K_s$. When $K_s$ is tuned to 19, the estimated number of clusters is equal to the ground-truth number of clusters of acoustic scenes (19 for LITIS-Rouen and 15 for DCASE-2017). In contrast, when $K_s$ deviates from 19, the estimated number of clusters is not equal to the ground-truth number of clusters of acoustic scenes. Hence, the value of $K_s$ is set to 19 in the experiments in the following subsections, in which the test data of LITIS-Rouen and DCASE-2017 are used to evaluate the performance of the proposed method.

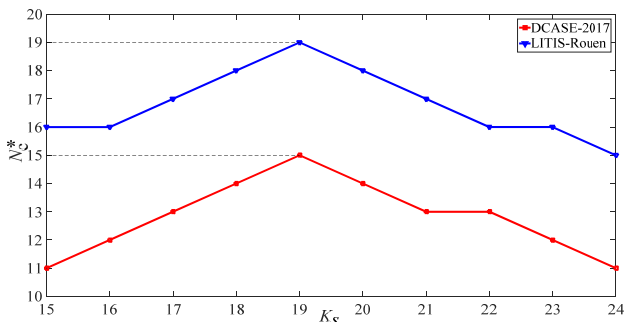

Fig. 5. Impacts of $K_s$ on the estimated optimal number of clusters $N_c^*$.

Fig. 6 presents the impacts of $N_c^*$ on the results of acoustic scene clustering when the proposed method is evaluated on LITIS-Rouen or DCASE-2017. We obtain the highest values of NMI and CA when $N_c^*$ is equal to the ground-truth number of clusters of acoustic scenes (19 for LITIS-Rouen and 15 for DCASE-2017). When $N_c^*$ deviates from the ground-truth number of clusters, the values of NMI and CA gradually decrease.

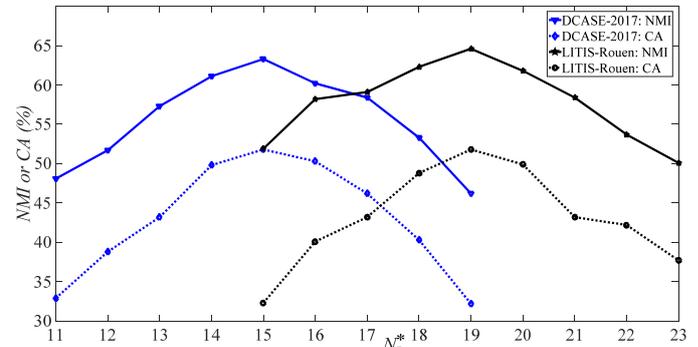

Fig. 6. Impacts of the estimated optimal number of clusters $N_c^*$ on the results of acoustic scene clustering.

### D. Comparison of Various Methods

In this subsection, we compare the proposed method with representative methods that were presented in previous works. The method that was proposed in [27] is the newest and most relevant method for acoustic scene clustering, in which the feature of MFCC and its first-order and second-order derivatives are fed into the algorithm of sparse subspace clustering (SSC) for acoustic scene clustering. The method that was presented in [28] is an agglomerative information bottleneck (AIB) algorithm that is fed by the MFCC feature. The methods in [29] and [30] are spectral clustering (SC) and AHC algorithms, respectively, and they are both fed by time-frequency features, such as MFCC, energy, and brightness. The K-means algorithm [40] is one of the most commonly used algorithms for data clustering and is frequently used as a baseline (fed by time-frequency features). These methods (including our method) are unsupervised methods. The baseline system of the DCASE 2017 challenge is also used as a baseline. It is a supervised method based on a multilayer perceptron architecture and uses log mel-band energies as input features [3]. The differences among these methods are summarized in Table VI.

TABLE VI
SUMMARY OF METHODS FOR ACOUSTIC SCENE CLUSTERING OR CLASSIFICATION

| Methods | Features | Classifier or clustering algorithm |
|---|---|---|
| Supervised [3] | Log mel energies | Multilayer perceptron |
| OURS | Deep embedding | Agglomerative hierarchical clustering |
| SSC [27] | MFCCs | Sparse subspace clustering |
| AIB [28] | MFCCs | Agglomerative information bottleneck |
| SC [29] | MFCCs, etc | Spectral clustering |
| AHC [30] | MFCCs, etc | Agglomerative hierarchical clustering |
| K-means [40] | MFCCs, etc | K-means |

The parameters of all methods are set according to the suggestions in the corresponding references and are optimally tuned on the development data. Then, all methods are evaluated on the test data. The class identities and numbers of classes of acoustic scenes are known in advance for the supervised method, whereas they are all unknown for the unsupervised methods. The values of the final affinity ratio ($\gamma^\tau$) are 2.85 and 2.13 for LITIS-Rouen and DCASE-2017, respectively.

Under the same conditions, the results that are obtained by the methods are listed in Table VII. The proposed method obtains *NMI* scores of 64.6% and 63.3% on the datasets of LITIS-Rouen and DCASE-2017, respectively. It realizes *CA* scores of 51.8% and 51.6% on the datasets of LITIS-Rouen and DCASE-2017, respectively. In terms of both *NMI* and *CA*, it outperforms other unsupervised methods, including the SSC [27], the AIB [28], the SC [29], the AHC [30] and the K-means [40] methods; however, it is outperformed by the supervised method [3]. The improvements of both *NMI* and *CA* that are realized by the supervised method are mainly due to the use of pretrained classifiers and the known number of classes of acoustic scenes. The main advantage of the proposed method over the supervised method is that it can obtain the number of clusters of acoustic scenes and can merge audio recordings of the same class of acoustic scene into a single cluster without using prior information. This advantage is essential for acoustic scene analysis on a huge mass of complex audio data, because prior information on the data is typically unknown and manual annotation of the data is expensive and time-consuming.

In terms of runtime, the proposed method is the most time-consuming. Both the updating of the CNN parameters for deep embedding extraction and the merging of clusters at each iteration are time-consuming. Hence, compared to the other methods, the proposed method requires more time to complete the same task. However, the runtime differences between the proposed method and the other methods are insignificant: the maximum differences are 1.5 (2.2 - 0.7) minutes on LITIS-Rouen and 1.6 (2.4 - 0.8) minutes on DCASE-2017.

TABLE VII
ACOUSTIC SCENE CLUSTERING/CLASSIFICATION PERFORMANCES OF VARIOUS METHODS (IN %)

| Methods | LITIS-Rouen | | | DCASE-2017 | | |
|---|---|---|---|---|---|---|
| | *NMI* | *CA* | RT (m) | *NMI* | *CA* | RT (m) |
| Supervised [3] | 69.3 | 55.3 | 0.7 | 68.2 | 55.1 | 0.8 |
| OURS | 64.6 | 51.8 | 2.2 | 63.3 | 51.6 | 2.4 |
| SSC [27] | 52.9 | 43.2 | 1.9 | 52.1 | 43.3 | 2.0 |
| AIB [28] | 59.3 | 48.2 | 1.4 | 58.1 | 46.8 | 1.5 |
| SC [29] | 56.2 | 45.1 | 1.8 | 56.4 | 45.2 | 2.0 |
| AHC [30] | 58.1 | 46.5 | 1.7 | 57.0 | 45.6 | 1.8 |
| K-means [40] | 54.8 | 44.7 | 1.2 | 54.3 | 44.1 | 1.5 |

RT (m): runtime in minutes.

The acoustic scene with the lowest values of both *NMI* and *CA* for the proposed method is *Pedestrian street* in LITIS-Rouen, and *Train* in DCASE-2017. In LITIS-Rouen, *Pedestrian street* is easily confused with other similar acoustic scenes, such as *Quiet street*, *Busy street* and *Market*. In DCASE-2017, *Train* tends to be confused with *Tram* and *Metro station*. These confusions that are caused by the proposed method are probably due to the large variations of time-frequency properties within these acoustic scenes and the overlaps in the distributions of the time-frequency properties of these acoustic scenes.

*E. Comparison of Features*

In this subsection, we compare the deep embedding (DE) with representative features that have been popularly adopted in previous works, such as the LMS [7], MFCCs [27]-[30], histogram of gradients (HOG) [12], I-vector [25], and bottleneck feature (BF) which is learned by a deep neural network [10]. The parameters for extracting these features are determined according to the suggestions in the corresponding references and optimally tuned on the experimental datasets. To compare the performances of these features under the same conditions, the same back-end clustering algorithm, namely, AHC, is used for all features in this experiment. The obtained results for various features are presented in Table VIII.

Table VIII
ACOUSTIC SCENE CLUSTERING PERFORMANCES OF VARIOUS FEATURES (IN %)

| Features | LITIS-Rouen | | DCASE-2017 | |
|---|---|---|---|---|
| | *NMI* | *CA* | *NMI* | *CA* |
| DE | 64.6 | 51.8 | 63.3 | 51.6 |
| LMS [7] | 55.1 | 44.7 | 54.9 | 44.8 |
| MFCCs [27]-[30] | 56.4 | 44.9 | 55.2 | 45.1 |
| HOG [12] | 57.5 | 45.3 | 55.8 | 46.2 |
| I-vector [25] | 57.0 | 45.1 | 56.1 | 46.7 |
| BF [10] | 58.2 | 46.5 | 57.3 | 47.8 |

According to Table VIII, when evaluated on LITIS-Rouen, feature DE obtains *NMI* of 64.6% and realizes gains of 9.5%, 8.2%, 7.1%, 7.6% and 6.4% over LMS, MFCCs, HOG, I-vector and BF, respectively. In terms of *CA*, the DE obtains 51.8%, and yields improvements of 7.1%, 6.9%, 6.5%, 6.7% and 5.3% over LMS, MFCCs, HOG, I-vector and BF, respectively. Similar results are obtained when these features are evaluated on the dataset of DCASE-2017. Based on these results, it is concluded that DE outperforms the other features in terms of both *NMI* and *CA*.

*F. Generalization across Clustering/Classification Algorithms*

In this subsection, we evaluate the generalizability of the learned deep embedding to other clustering/classification techniques. The features that are used in the various methods are all replaced with the deep embedding, which is learned via the proposed framework for acoustic scene clustering, and the parameters of the clustering/classification algorithms are not changed. Table IX presents the results that were obtained by the methods after using the learned deep embeddings as input features. All clustering/classification algorithms (except ours) realize higher values of *NMI* and *CA* than their counterparts in Table VII and the variances in these two metrics across all clustering algorithms are small. The AHC method of [30] is the same as ours if the deep embedding is used as its input feature. In addition, the values of *NMI* and *CA* that are obtained by the supervised method are still much higher than those that are obtained by all the unsupervised methods. In conclusion, the deep embedding generalizes well across various clustering algorithms instead of overfitting a single clustering algorithm.

Table IX
ACOUSTIC SCENE CLUSTERING/CLASSIFICATION PERFORMANCES OF VARIOUS METHODS IN WHICH DEEP EMBEDDINGS (DES) ARE USED AS INPUT FEATURES (IN %)

| Methods | LITIS-Rouen | | DCASE-2017 | |
|---|---|---|---|---|
| | NMI | CA | NMI | CA |
| DE + Supervised [3] | 75.1 | 60.4 | 73.6 | 59.4 |
| Ours | 64.6 | 51.8 | 63.3 | 51.6 |
| DE + SSC [27] | 59.3 | 48.8 | 58.2 | 47.5 |
| DE + AIB [28] | 61.1 | 50.3 | 60.2 | 49.2 |
| DE + SC [29] | 60.3 | 50.1 | 59.3 | 48.4 |
| DE + AHC [30] | 64.6 | 51.8 | 63.3 | 51.6 |
| DE + K-means [40] | 58.6 | 46.7 | 57.1 | 46.2 |

*G. Generalization across Corpora*

In this subsection, we evaluate the generalizability of the deep embedding across corpora. We construct a CNN that is based on the proposed method on one corpus (the source dataset). Then, we extract deep embeddings from the other corpus (the target dataset) using the preconstructed CNN. Finally, we conduct acoustic scene clustering on the deep embeddings via the AHC algorithm. In the first column of Table X, the item on the left side of the arrow (e.g., LITIS-Rouen in "LITIS-Rouen → DCASE-2017") denotes the source dataset, while that on the right side of the arrow (e.g., DCASE-2017 in "LITIS-Rouen → DCASE-2017") denotes the target dataset. The experimental results are listed in Table X. The deep embedding realizes superior performance when the source dataset and the target dataset are from the same corpus instead of from different corpora. However, the absolute values of the performance differences between these two conditions (source and target datasets from the same versus different corpora) do not exceed 4.4%. For example, the differences of *NMI* and *CA* between the condition LITIS-Rouen → LITIS-Rouen and the condition LITIS-Rouen → DCASE-2017 are 4.4% and 2.5%, respectively. Hence, the learned deep embedding can be well transferred across corpora.

Table X
ACOUSTIC SCENE CLUSTERING PERFORMANCE OF DEEP EMBEDDING ACROSS CORPORA (IN %)

| | NMI | CA |
|---|---|---|
| LITIS-Rouen → DCASE-2017 | 60.2 | 49.3 |
| LITIS-Rouen → LITIS-Rouen | 64.6 | 51.8 |
| DCASE-2017 → LITIS-Rouen | 61.1 | 50.1 |
| DCASE-2017 → DCASE-2017 | 63.3 | 51.6 |

*H. Performance on Imbalanced Data*

In this subsection, we evaluate the performance of the proposed method on imbalanced data. We sample subsets of LITIS-Rouen and DCASE-2017 with various retention rates. Audio recordings of the first class of acoustic scenes will be kept with the minimum retention rate $R_{min}$. That is, $R_{min}$ times of audio recordings of the first class of acoustic scenes are used in experiments. All audio recordings of the last class of acoustic scenes are used in the experiments. Audio recordings of other classes (the second to the eighteenth classes for LITIS-Rouen and the second to the fourteenth classes for DCASE-2017) are retained at rates that vary linearly in between $R_{min}$ and the maximum rate of 1. As a result, the largest class is $1/R_{min}$ times as large as the smallest class. The process of class selection is repeated 10 times, where both the first and last classes are randomly selected and differ each time. The results that are obtained by the proposed method are presented in Table XI, where $R_{min}$ is set to 0.1. The number of clusters that is estimated by the proposed method is equal to the ground-truth number of classes of acoustic scenes (19 for LITIS-Rouen and 15 for DCASE-2017) in 6 of the 10 cases and 5 of the 10 cases for the LITIS-Rouen and the DCASE-2017, respectively. In addition, the values of *NMI* and *CA* are similar among cases (either correct or incorrect estimation of the number of clusters) when the proposed method is evaluated on the two corpora. That is, the proposed method can correctly estimate the number of clusters most of the time with higher values of metrics on imbalanced data. In conclusion, the proposed method is robust against class size variation.

TABLE XI
INFLUENCE OF IMBALANCED DATA ON THE ACOUSTIC SCENE CLUSTERING PERFORMANCE OF THE PROPOSED METHOD (IN %)

| Times | LITIS-Rouen | | | DCASE-2017 | | |
|---|---|---|---|---|---|---|
| | NMI | CA | ECN | NMI | CA | ECN |
| 1 | 64.6 | 51.8 | 19 | 63.3 | 51.6 | 15 |
| 2 | 61.5 | 50.8 | 19 | 60.6 | 49.2 | 14 |
| 3 | 55.8 | 43.1 | 17 | 57.2 | 46.8 | 14 |
| 4 | 62.0 | 50.5 | 19 | 54.5 | 42.4 | 13 |
| 5 | 60.2 | 49.9 | 19 | 62.2 | 50.4 | 15 |
| 6 | 54.3 | 42.2 | 17 | 62.8 | 50.6 | 15 |
| 7 | 61.8 | 50.6 | 18 | 54.4 | 42.4 | 13 |
| 8 | 60.4 | 48.1 | 19 | 61.7 | 49.2 | 15 |
| 9 | 60.8 | 50.2 | 19 | 56.9 | 44.1 | 14 |
| 10 | 59.6 | 49.6 | 18 | 59.3 | 49.7 | 15 |
| Mean | 60.1 | 48.7 | 18.4 | 59.3 | 47.6 | 14.3 |

ECN: estimated class number.

## VI. CONCLUSIONS

We have addressed acoustic scene clustering without using manual labels by jointly optimizing the procedures of deep embedding learning and agglomerative hierarchical clustering iteration. The joint optimization of these procedures is directed by a unified loss function. The graph degree linkage is introduced for measuring the similarities between pairs of clusters, while an affinity ratio is defined for determining the optimal number of clusters.

Based on the presentation of the proposed method and the experimental results, the following conclusions are drawn.
1) The proposed method is an unsupervised method that does not use any prior information (e.g., manual labels and numbers of classes) on the audio recordings in advance. It can be used to cluster a huge mass of audio recordings. In addition, it outperforms state-of-the-art unsupervised methods in terms of both *NMI* and *CA* and remains effective on imbalanced data.
2) The deep embedding feature that is learned by the proposed framework can capture the characteristics of various acoustic scenes and can be used as an effective feature for clustering acoustic scenes. It outperforms the features that were popularly used in previous works, such as LMS, MFCCs, HOG, I-vector, and BF. In addition, it generalizes well across various clustering algorithms and across corpora.

Our future work will consist of three parts. First, we will consider other types of networks for learning deep embeddings and other clustering algorithms in the proposed joint optimization framework. Second, we will explore other methods for measuring the affinities between clusters and for estimating the optimal number of classes during clustering procedures. Finally, we will explore how the generalizability of the proposed method is influenced by the value of integer $K_s$. For example, what if the ground-truth label is not available, namely, no development dataset is available? What if there are multiple $K_s$ values that lead to the same $N_c^*$ value? If no values in the search range of $K_s$ correspond to $N_c^*$ in the development dataset, what value should be selected?